\documentclass[12pt]{article}
\usepackage{graphicx}
\newcommand{\beq}{\begin{equation}}
\newcommand{\eeq}{\end{equation}}
\newcommand{\bea}{\begin{eqnarray}}
\newcommand{\eea}{\end{eqnarray}}

\newcommand{\m}{\mu}

\begin{document}
\begin{flushleft}
       \hfill                      {\tt hep-th/0212102}\\
       \hfill                       FIT HE - 02-07 \\
       \hfill                       KYUSHU-HET 63 \\
\end{flushleft}


\begin{center}
{\bf\LARGE Gauge symmetry and cosmological constant on a brane \\ }
\vspace*{12mm}
{\large Kazuo Ghoroku\footnote[2]{\tt gouroku@dontaku.fit.ac.jp} and
Nobuhiro Uekusa\footnote[3]{\tt uekusa@higgs.phys.kyushu-u.ac.jp}}\\
\vspace*{2mm}

\vspace*{2mm}

\vspace*{4mm}
{\large ${}^{\dagger}$Fukuoka Institute of Technology, Wajiro, Higashi-ku}\\
{\large Fukuoka 811-0295, Japan\\}
\vspace*{4mm}
{\large ${}^{\ddagger}$Department of Physics, Kyushu University, Hakozaki, 
Higashi-ku, Fukuoka 812-8581, Japan}\\
\vspace*{10mm}
\end{center}




\begin{abstract}
We have examined the localization of gauge bosons on the three-brane 
embedded in 5D bulk space-time, and we find two kinds of branes on
which both graviton and gauge bosons can be trapped. 
One is the dS brane with a positive cosmological constant, and the
other is the one with zero cosmological constant or 4D Minkowski brane.
Then,
which brane is realized would depend on the observation of the cosmological
constant of our world.

\end{abstract}


\newpage

\section{Introduction}

It is fascinating to regard our four dimensional world 
as a three-brane ~\cite{RS1,RS2} embedded in a 
higher dimensional space with extended codimensions. 
In the recent progress of superstring theory
or M-theory, we could find an appropriate non-perturbative solution
which could be connected to a realistic brane-world.
However, there are many things to be cleared before arriving at such a
solution.

The idea of three-brane involves many interesting aspects
to approach several fundamental problems. Actually, it
opened an alternative to the standard Kaluza-Klein (KK)
compactification via the localization of the graviton 
in the 4D Minkowski space \cite{RS2} and also in the 4D de Sitter space
\cite{bre,BG,NO}. In order to confirm the idea of braneworld,
it would be important to see the localization of the
gauge bosons, which is an important force to construct our
world, i.e. the Coulomb force. 
However, gauge bosons feel the warp factor in the bulk differently
from the case of the graviton, and it is
a difficult problem how to localize the gauge bosons on the brane.

In the case of RS model, where 4d cosmological constant $\lambda$ is zero, some
ideas for the localization of gauge bosons have been proposed (for example,
see \cite{KT,GN2}). In general,
we can expect that the situation of this problem would be changed when a small
$\lambda$ is considered on the brane since the warp factor is changed 
in this case compared to the naive model. Actually, we could find
the localization of a massive 
mode of a scalar in this case \cite{GY2}.

Our purpose is to extend this analysis to the case of vector fields to
see the localization of bulk vector-fields as gauge
bosons on the brane. The localization of graviton or
massless scalar can not give any restriction to the magnitude of the
$\lambda$, and we can see the localization of these fields in a wide
range of the parameters in the theory. While, 
we point out that localization of gauge bosons intimately is related to
the value of $\lambda$ as shown in the analyses given here.

For a brane with a positive $\lambda$,
the bulk gauge bosons are trapped on it. And we should notice
the recent
observation of small inflation of our universe
which implies the existence of small positive $\lambda$. 

At the same time, we propose another kind of brane proposed previously 
by one of us \cite{GN2}. In this case, the bulk vector is massive
and a mass term of the vector is also attached on the brane action.
The origin of the latter term might be found as quantum corrections 
in the bulk \cite{GGH}. In this case, we could find a gauge boson
trapped on the brane with $\lambda=0$
when we choose an appropriate relation between
the bulk and brane-coupled mass terms.

\vspace{.5cm}
In Section 2, we give our model in the bulk and 
a brief review of brane solutions used here.
In Section 3, for the case including manifest bulk gauge symmetry,
the localization of the massive and massless 
vectors on the brane is examined. 
In Section 4, for the case with the brane coupling, the vector localization 
is examined. 
Concluding remarks are given in the final section.

\section{Brane model and solutions with cosmological constant}

Here we set up our model by the following effective action, 
\beq
 S= S_{\rm gr}+S_{\rm A}.
\eeq
The first term denotes the gravitational part, 
\beq
    S_{\rm gr} = {1\over 2\kappa^2}\Bigg\{
      \int d^5X\sqrt{-G} (R -2 \Lambda)
          +2\int d^4x\sqrt{-g}K\Bigg\}-{\tau}\int d^4x\sqrt{-g}, 
                  \label{action}
\eeq
where $K$ is the extrinsic curvature on the boundary, and $\tau$ represents 
the tension of the brane which is situated at $y=0$
by imposing $y\leftrightarrow -y$ symmetry for action.
The second part $S_{\rm A}$ denotes the action for the massive vector, 
which is denoted by $A_M(x,y)$, ($M=0,\cdots, 4$), 
\beq
 S_{\rm A} = \int d^5X\sqrt{-G}\Bigg\{-{1\over 4}G^{MN}G^{PQ}F_{MP}F_{NQ}
               - {1\over 2}(M^2+c~\delta(y)) G^{MN}A_M A_N\Bigg\} \, ,\label{SA}
\eeq
where the parameter $M$ and $c$ denote the bulk mass of the vector and 
the coupling of the vector to the brane respectively.  
In the case of Randall-Sundrum solution with zero-cosmological constant, 
the massive vector can be trapped as a photon 
only
when we take the parameter $c$
as follows \cite{GN2},
\beq
 c=-2\mu(\sqrt{1+{M^2\over \mu^2}}-1)\equiv\hat{c},  \label{cval}
\eeq
where $\mu=\sqrt{-\Lambda/6}$.
This relation
seems to be a kind of fine-tuning of the parameters in the theory. 
Here we extend the analysis to the case of $\lambda\ge 0$ to see 
the localization of vector.

\vspace{.5cm}
We consider the vector-field fluctuations on the background given as a
solution of (\ref{action}).
Here, the background configuration with non-zero $\lambda$ 
is considered, and
the Einstein equation is solved in the following metric,
\beq
 ds^2= A^2(y)\left\{-dt^2+a^2(t)\gamma_{ij}(x^i)dx^{i}dx^{j}\right\}
           +dy^2  \, \label{metrica},
\eeq
where the coordinates parallel to the brane are denoted by $x^{\mu}=(t,x^i)$,
and $y$ being the coordinate transverse to the brane. 
We restrict our interest here to the case of a
Friedmann-Robertson-Walker type
universe. Then, the three-dimensional metric $\gamma_{ij}$
is described in Cartesian coordinates as
\beq
  \gamma_{ij}=(1+k\delta_{mn}x^mx^n/4)^{-2}\delta_{ij},  \label{3metric}  
\eeq
where the
curvature indices 
$k=0, 1, -1$ correspond to a
flat, closed, or open universe respectively. In this brane-world, the 
effective 
cosmological constant $\lambda$ on the brane is given as
\beq
   \lambda = \kappa^4\tau^2/36+\Lambda/6 . \label{4cos}
\eeq 
We notice that $\lambda$ depends on both parameters $\tau$ and $\Lambda$. 
The solutions of non-zero $\lambda$ used here are summarized below.

\vspace{.3cm}
\underline{\bf (I) $\lambda >0$}:  
When $\lambda >0$, we obtain the solution for 
$a_0(t)$ which represents inflation of three 
dimensional
space. Here we give the
simple one, which is used hereafter, for the case of $k=0$,
\beq
 a_0(t)=e^{H_0t},
\eeq
where the Hubble constant is represented as $H_0=\sqrt{\lambda}$.
We notice that $a_0(t)$ has nothing to do with the problem of
localization and it depends only on the form of $A(y)$ as we will see
below. 

\vspace{.2cm}
As for $A(y)$, we obtain the same solution for any value of $k=0, \pm1$. 
The solution of $A(y)$ depends on the sign of $\Lambda$
even if the same positive $\lambda$ is assigned.

\underline{\bf (I-1) $\Lambda<0$}: 
For $\Lambda <0$, $A(y)$ is solved as
\beq
  A(y)= {\sqrt{\lambda}\over\mu} \sinh[\mu(y_H-|y|)] \, \label{metrica4}
\eeq
\beq
  \qquad \sinh(\mu y_H)=\mu/\sqrt{\lambda}.
               \label{constads}
\eeq
where $y_H$ represents the position of the horizon in the bulk.
This solution and others are normalized at $y=0$, as $A(0)=1$. 
And all the background configurations are taken
to be $Z_2$ symmetric with respect to the reflection, $y\to -y$. 

\underline{\bf (I-2) $\Lambda>0$}: 
When  $\Lambda$ is positive, the solution for  
 $a_0(t)$ is the same as above, but  
 $A(y)$ is different.
One has
\beq
 A(y)={\sqrt{\lambda}\over \mu_d}\sin[\mu_d(y_H-|y|)],
          \label{desit}
\eeq 
\beq
  \mu_d=\sqrt{\Lambda/6}, \qquad \sin(\mu_d y_H)=\mu_d/\sqrt{\lambda}.
               \label{const1}
\eeq
These two configurations represent $dS_4$-branes embedded in the
bulk at $y=0$.

\vspace{.2cm}

\underline{\bf (II) $\lambda <0$}: When $\lambda<0$, we obtain 
\beq
 a_{-1}(t)={1\over H} \sin(H t), \quad H=\sqrt{-\lambda}
 \label{a0-ads4}
\eeq
and $k=-1$. 
The $AdS_4$-brane is obviously obtained for $\Lambda <0$ because of 
(\ref{4cos}). And $A(y)$ is given as

\beq
 A(y)= {H\over\mu} \cosh[\mu(y_{H}-|y|)] \, \label{metrica-AdS4}
\eeq
\beq
  \cosh(\mu y_{H})=\mu/H .
               \label{const-AdS4}
\eeq
Note that the bulk has no horizon in this case in spite of the letter
$y_H$ used above. 
$AdS_4$ brane with $\lambda<0$ can not be regarded
as our world from the viewpoint of brane-world
since the general coordinate invariance is lost \cite{KR}.
However, we consider this case from the theoretical viewpoint.

\section{Vector localization: $c=0$ case}

Here we concentrate our attention on the behaviors of the 
vector-field fluctuation around the background (\ref{metrica}) 
by extending the previous analysis given in \cite{GN2}. 

\vspace{.3cm}
At first, we examine this problem for the case of $c=0$ in the action
(\ref{SA}). It contains the case where bulk gauge symmetry is 
manifest as a limit of $M=0$.
The field equation of $A_M$ is given as, 
\beq
 {1\over\sqrt{-G}}\partial_A[\sqrt{-G}
  (G^{AB}G^{CD}-G^{AC}G^{BD})\partial_B A_C] -
           M^2 G^{DB}A_B =0 . \label{eq100}
\eeq
This equation can be solved by using the identity obtained 
in the massive 
vector theory,
\beq
  \partial_A(\sqrt{-G}G^{AB}A_B)=0, \label{identity300} 
\eeq
where we should notice that for $M=0$,
we interpret Eq.(\ref{identity300}) as the gauge condition.
We solve these equations by rewriting the five components of $A_M$ as 
($A_t$, $A_i^T$, $A_i^L$, $A_y$), where the 3D transverse part is defined
as $\nabla_i A^i_T=0$ and $A_i^L$ denotes the longitudinal component.

\vspace{.5cm}
The next step is to expand the fields in terms of the four-dimensional 
mass eigenstates:
\beq
 A_M(x,y)=\int dm ~~\tilde{a}_M(m,t,x^i)\phi_M(m,y) \, , \label{eigenex}
\eeq
where the mass $m$ is defined for each $\tilde{a}_M(m,t,x^i)$ as shown below.
For $A_i^T$, we obtain,
\beq
  (\partial_y^2+2{A'\over A}\partial_y+{m^2\over A^2})\phi_i^T
      =M^2 \phi_{i}^{T} , 
\label{tra00}
\eeq
\beq
  (-\partial_t^2-{\dot{a}_0\over a_0}\partial_t+{\partial_i^2\over a_0^2})
   \tilde{a}_i^T
       =m^2 \tilde{a}_i^T ,  
\label{trat00}
\eeq
where ${}'=d{}/dy$ and $\dot{}=d{}/dt$. In deriving the above two equations,
Eq.(\ref{identity300}) is not used.
Since Eq.(\ref{identity300}) 
give a relation among $A_y$, $A_{\mu}^L$ and $A_t$,
it is used below for the equations for those components.
The second equation (\ref{trat00}) gives a definition of the 4D mass 
$m$ for the transverse
components, and this definition is not common to other components, which are
shown below. However we notice that they coincide when $a_0=1$.
The equations for other components can be written as follows:

\vspace{.5cm}
For $A_y$,
\beq
  (\partial_y^2+6{A'\over A}\partial_y+4[{A''\over A}+({A'\over A})^2]
      +{m^2\over A^2})\phi_y
      =M^2 \phi_y , 
\eeq
\beq
  (-\partial_t^2-3{\dot{a}_0\over a_0}\partial_t+{\partial_i^2\over a_0^2})
   \tilde{a}_y
       =m^2 \tilde{a}_y ,  
\eeq

\vspace{.5cm}
For $A_t$,
\beq
  (\partial_y^2+2{A'\over A}\partial_y+{m^2\over A^2})\phi_t
      =M^2 \phi_{t}-{1\over \tilde{a}_t}O_t\tilde{a}_y\phi_y ,
\eeq
\beq
  (-\partial_t^2-5{\dot{a}_0\over a_0}\partial_t+{\partial_i^2\over a_0^2}
      -3[({\dot{a}_0\over a_0})^2+{\ddot{a}_0\over a_0}])\tilde{a}_t
       =m^2 \tilde{a}_t ,  
\eeq
\beq
 O_t=2\left\{{\dot{a}_0\over a_0}A(4A'+A\partial_y)+A'A\partial_t\right\}
\eeq

\vspace{.5cm}
For $A_i^L$,
\beq
  (\partial_y^2+2{A'\over A}\partial_y+{m^2\over A^2})\phi_i^L
      =M^2 \phi_{i}^{L}
    -{2\over \tilde{a}_i^L}\partial_i({A'\over A}\tilde{a}_y\phi_y
       -{\dot{a}_0\over a_0A^2}\tilde{a}_t\phi_t) , 
\eeq
\beq
  (-\partial_t^2-{\dot{a}_0\over a_0}\partial_t+{\partial_i^2\over a_0^2})
   \tilde{a}_i^L
       =m^2 \tilde{a}_i^L .   \label{transv-t}
\eeq
Here, we comment on the $k$-dependence of the above equations, which are
written for $k=0$ of solutions {\bf (I)}.
We consider also the solution 
{\bf (II)} in which $k=-1$. 
It can be seen that the equations for $\phi_M(y)$
are not changed but $\partial_i$ is changed by the three dimensional
covariant derivative $\nabla_i(\gamma_{lm})$ in the other equations.
And this changing gives no effect on the following analyses.

\vspace{.5cm}

As is well known, the localization is determined by the equations of 
$\phi_M(y)$, and we do not need to know the details of $\tilde{a}_M(t,x^i)$. 
Although the equations for $\phi_t(y)$ and $\phi_i^L(y)$ include the
inhomogeneous part, the homogeneous part of equations are 
equivalent to the one of $\phi_i^T$. We can solve these equations
for $\phi_t(y)$ and $\phi_i^L(y)$ by separating them into the
special solutions, which satisfy the
inhomogeneous equations, and the solutions, which satisfy
the same form equation of (\ref{tra00}), the homogeneous part.
Then the general solutions are written as
\beq
    \phi=\phi_{\rm special} +\phi_{i}^T.
\eeq
We expect that the localized modes are included in this general solutions
through $\phi_{i}^T$ if $A_i^T$ were trapped.
Then the equations
for $\phi_M(y)$ are separated to two groups of $A_{\mu}$=($A_t$,$A_i$) 
and $A_y$ from the viewpoint of localization.

\vspace{.3cm}
However
$A_y$ is an odd function with respect to $y$ 
so we can say that $A_y|_{y=0}=0$,
then this component is not localized on the brane. While, we can see that
the other components $A_{\mu}$, which are parallel to the brane, 
are even functions of $y$ from Eq.(\ref{identity300}) 
which is written as,
\beq
  \partial_A(\sqrt{-G}G^{AB}A_B)={1\over A^2}\partial_y(A^4A_y)-{1\over a_0^3}\partial_t(a_0^3A_t)
      +{1\over a_0^3}\partial_i(a_0A_i),  \label{identity4}
\eeq
where we notice that the warp factor $A(y)$ is an even function of $y$. 
Then $A_{\mu}$
could be localized on the brane, and it is enough to examine only
$A_i^T$ for the problem of localization of the gauge boson since 
the homogeneous part of the equations of other components 
obeys the common equation with the one of $A_i^T$ as stated above.
So we concentrate on $A_i^T$ hereafter.

\vspace{.5cm}

Consider the equation (\ref{tra00}) for $A_i^T$.
For any solution of the background metric, $A(y)$ and $a_0(t)$,
Eq.(\ref{tra00}) can be rewritten in a one dimensional Schr\"{o}dinger-type
equation 
\beq
 [-\partial_z^2+V(z)]u(z)=m^2 u(z) , \ \label{warp300}
\eeq
by introducing $u(z)$ and $z$ defined as 
$\phi_i^T=A^{-1/2}u(z)$ and $\partial z/\partial y=\pm A^{-1}$.
Here $A(y)$ is normalized as $A(0)=1$ and the potential $V(z)$ is given as
\beq
 V(z)={1\over 4}(A')^2+{1\over 2}AA''+A^2M^2 ,  \label{pott00}
\eeq 
where $z_0$ represents the value of $z$ at $y=0$ and it depends
on the solutions, $A(y)$, as well as $V(z)$. Here we should notice,
$A'=\partial A/\partial y$ and the metric form 
(\ref{metrica}).\footnote[5]{ 
When we use the notation, $A'=\partial A/\partial z$ and 
$
 ds^2= e^{2\tilde{A}}\left\{-dt^2+a^2(t)\gamma_{ij}(x^i)dx^{i}dx^{j}\right\}
           +dy^2
$, we obtain $V(z)={1\over 4}(\tilde{A}')^2+{1\over 2}\tilde{A}''
+e^{2\tilde{A}}M^2$. This form would be more popular for some people.
}

\vspace{.5cm}
The potentials are obtained for the three background solutions given in the
previous section as follows,

{\bf (I-1) Solution for $\lambda>0$ and $\Lambda<0$ :}

\beq
 V(z)= {1\over 4}\lambda [{4M^2/\mu^2+3\over\rm{sinh}^2(\sqrt{\lambda}|z|)}+1]
      -{\kappa^2\tau\over 6}\delta(|z|-z_0) , \label{pot300}
\eeq
$$z=\rm{sgn}({\it y})(\lambda)^{-1/2}\ln(\coth[\mu ({\it y}_H-|{\it y}|)/2])
$$
\beq
 z_0={1\over\sqrt{\lambda}}\rm{arcsinh}({\sqrt{\lambda}\over \mu}).
\eeq

{\bf (I-2) Solution for $\lambda>0$ and $\Lambda>0$ :} 

\beq
 V(z)= {1\over 4}\lambda [{4M^2/\mu_d^2-3\over\rm{cosh}^2(\sqrt{\lambda}z)}+1]
      -{\kappa^2\tau\over 6}\delta(|z|-z_0) , \label{pot200}
\eeq
$$z=\rm{sgn}({\it y})(\lambda)^{-1/2}\ln(\cot[\mu_d({\it y}_H-|{\it y}|)/2])$$
\beq
 z_0={1\over\sqrt{\lambda}}\rm{arccosh}({\sqrt{\lambda}\over \mu_d}).
\eeq

{\bf (II) Solution for $\lambda<0$ and $\Lambda<0$ :}

\bea
  &&V(z)= {1\over 4}H^2\left[{4M^2/\mu^2+3\over\cos^2(H z)}-1\right]
      -{\kappa^2\tau\over 6}\delta(|z|-z_0) \ , \label{pot400}
\\
  &&z=2\rm{sgn}({\it y})\arctan\tanh\left[\mu(y_H-|y|)/2\right])/H \ ,\nonumber
\\
  &&\qquad\qquad\qquad
    z_0={1\over H}\arccos({H \over \mu}) \ .
\eea
Before solving the above equation with the each potential obtained here,
we should notice that only if the eigenvalue $m^2$ is smaller than the minimum
of the potential except the $\delta$ function term which is denoted as
$V_{\textrm{\scriptsize min}}$,
the corresponding field has the possibility to be trapped.
For {\bf (I)}, this minimum value is 
$V_{\textrm{\scriptsize min}}=\lambda/4$.
On the other hand for {\bf (II)}, the minimum is given by
$V_{\textrm{\scriptsize min}}=M^2+(3\mu^2-H^2)/4$.
The fluctuation with the mass $m^2>V_{\textrm{\scriptsize min}}$
corresponds to continuum mode.

\vspace{.5cm}
\noindent{\bf Solution(I-1)}
First, we consider the case {\bf (I-1)} 
where $\lambda=0$ is realized when a
fine-tuning condition between $\kappa$ and $\Lambda<0$ is satisfied 
as in the RS solution.
The RS solution would be obtained
naturally when conformal invariance of the bulk is preserved for the
brane-world solution \cite{GY}. Here this symmetry is not clear, so it will be
natural to consider the general case of $\lambda\neq 0$.
Among them, we firstly consider the case of $\lambda>0$, {\bf (I-1)}.

In this background, $u$ is solved in terms of the following
general solution, which is given by ignoring the $\delta$-function
potential,
\beq
 u(z)=c_1X^{-id} {}_2F_1(b_1,b_2;\bar{c};-X)
+c_2X^{id}{}_2F_1(b_1',b_2';\bar{c}';-X), 
        \label{solds00}
\eeq
where $c_1$ and $c_2$ are constants of integration and
\beq
 X={1\over\rm{sinh}^2(\sqrt{\lambda}z)}, \quad 
      d={\sqrt{-1+4m^2/\lambda}\over 4},   \label{para100}
\eeq
\beq
  b_1={1\over 4}(1-\sqrt{1+\hat{a}})-id, 
   \quad b_2={1\over 4}(1+\sqrt{1+\hat{a}})-id, 
   \quad\bar{c}=1-2id, \label{para200}
\eeq
\beq
  b_1'={1\over 4}(1-\sqrt{1+\hat{a}})+id, 
     \quad b_2'={1\over 4}(1+\sqrt{1+\hat{a}})+id, 
     \quad \bar{c}'=1+2id. \label{para300}
\eeq
\beq
  \hat{a}=4M^2/\mu^2+3 .
\eeq
Here ${}_2F_1(\alpha,\beta;\gamma;x)$ 
denotes the Gauss's hypergeometric function.
It follows from this solution that $u(z)$ oscillates 
with $z$ when $m^2>{\lambda}/4$, where the continuum KK modes appear. 

\vspace{.5cm}
Here we concentrate on the bound state which is restricted to the region
of $m^2<{\lambda}/4$. 
We note also $V_{\textrm{\scriptsize min}}=\lambda/4$.
In this case,
$d$ in (\ref{para100}) is rewritten as $d=i\bar{d}$, where $\bar{d}$
is real and given below. We should notice this difference from the
solution given in the KK mode region, and we need a converging solution
in the region of the expected bound state.
So the solution is obtained
by setting $c_2=0$ and it is written as,
\beq
 u(z)=c_1X^{\overline{d}} {}_2F_1(b_1,b_2;\bar{c};-X),\qquad
    \overline{d}={\sqrt{1-4m^2/\lambda}\over 4} \ ,
 \label{solb00}
\eeq
which tends to zero in the limit of $z=\infty$ as $u(z)\to X^{\sqrt{1-4m^2/\lambda}/4}$. Then this solution is normalizable when $m^2<\lambda/4$. 
One should notice here that we imposed one boundary condition 
at $z=\infty$ on the solution,
which is written by two independent solutions.

The boundary condition at $z=z_0$ is given as follows
by taking into account of the 
$\delta$-function potential in (\ref{pot300}),
\beq
{du(z_0)\over dz}=-{\kappa^2\tau\over 12}u(z_0).  \label{boundzero00}
\eeq
\\

\noindent From this setting we find the following results;

\vspace{.3cm}
\underline{{\bf (1)} Normalizability}: 
The necessary condition of the localized state
is the normalizability of this state, then the integration over $y$ of the
effective action for this mode should be finite. It is easy to see that
this is equivalent to demand the following condition for $u(z)$,
\beq
  \int_{z_0}^{\infty} dz~u(z)^2 < \infty .  \label{norm}
\eeq
As stated above this condition implies $m^2<\lambda/4$ for our solution.
We should notice that 
the solution for $M^2=0$ is excluded
in the case of $\lambda=0$
since here the solution is a constant.
But this is not necessarily true in the case of $\lambda>0$ 
since the solution is not a constant even if $M^2=0$.
And
the condition $m^2<\lambda/4$ is a very loose condition which is
equivalent to the condition of non KK-mode.
The zero-mode, $m=0$, of course
satisfies this condition for $\lambda>0$ even if
how small it is. This fact is
important since we don't need any bulk mass $M$ for the vector field to 
obtain the normalizable zero-mode as in the case of RS brane of
$\lambda=0$ 
\cite{GN2,BG} 
. In other words, any gauge bosons are trapped on the 
brane in a gauge invariant form when a positive cosmological constant 
$\lambda$ exist.

In fact,
we can see below that the zero mode eigenstate satisfies also the
boundary condition (\ref{boundzero00}) which is needed from 
the equation (\ref{warp300}) for
the precise form of potential (\ref{pot300}).
 
\vspace{.5cm}

\underline{{\bf (2)} 
Relation between $m^2$ and $M^2$}: 
By solving (\ref{boundzero00}), we find the $M$-dependence of the
lowest eigenvalue $m$ near $M^2=0$. It behaves as
\beq
 m^2=\alpha M^2,    \label{alph}
\eeq
where $\alpha$ depends on the other parameters $\lambda$ and $\Lambda$.
Here we can see $\alpha>0$.
This is confirmed both in numerically and analytically. 

\begin{figure}[htbp]
\begin{center}
\voffset=15cm
  \includegraphics[width=8.5cm,height=7cm]{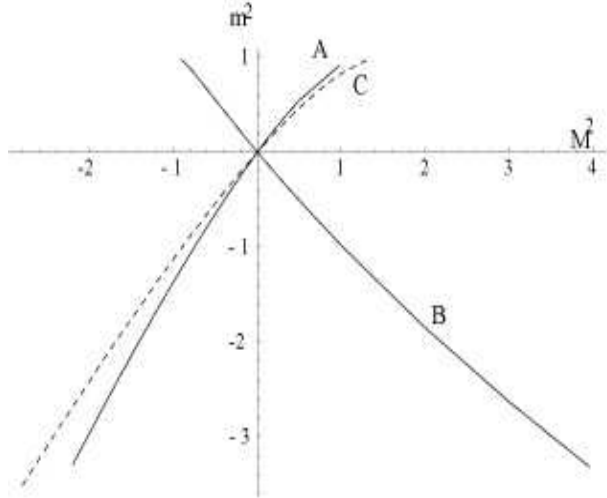}
\caption{The solid curves A and B show the eigenvalues $m^2$ versus $M^2$ 
for solutions {\bf (I)} for $c=0$ and $c=\hat{c}$ respectively, 
where $\lambda=\mu=1$. The $m^2$-$M^2$ curve for 
solution {\bf (II)} is shown by dashed curve for $\lambda=1.5,~ 
\mu=1$. In both cases, $m^2$ is scaled by $\lambda/4$.
   \label{mMfig}}
\end{center}
\end{figure}

Firstly, we consider the eigenvalue $m^2$ analytically. In general, 
$m^2$ can be expressed as
\beq
  m^2=\int dz u^{*}(z)[-\partial_z^2+V(z)]u(z)\equiv
       \langle u|[-\partial_z^2+V(z)]|u\rangle ,  \label{masseig3}
\eeq
where $u(z)$ denotes the normalized eigenfunction, $\langle u|u\rangle =1$.
Then we expand $m^2$ as 
\beq
   m^2=m_0^2+\alpha M^2  \label{mMrela}
\eeq
for small $M^2$, and $\alpha$ is expressed as
\beq
  \alpha=\partial_{M^2}m^2|_{M^2=0}=\langle u|A^2|u\rangle \ .
        \label{massei400}
\eeq
So we can see $\alpha>0$.
These points are assured by expanding the solution for $m^2$ obtained
from Eq.(\ref{boundzero00}) near an appropriate point. Here we examined
$m^2$ near small $\lambda$ and the following is found,
\beq
 m^2={1\over 2}M^2 \ .   \label{massei5}
\eeq
This result
implies $m_0=0$ for the solution {\bf (I-1)}. 
Therefore we can say that zero-mode of the vector field is localized when
its bulk mass $M$ is zero.
Then, all the gauge bosons without any
gauge-symmetry breaking in the bulk are localized on the brane.
This is completely different from the RS brane.
And the massive bulk-vectors ($M^2>0$) are trapped as 
massive vectors ($m^2>0$) 
on the brane. In this sense,
the situation is 
similar to the case of the graviton and scalar
fields.
In other words, the general coordinate invariance in the bulk
is also preserved on the dS brane as well as the gauge symmetry.
The essential point here is the presence of positive $\lambda$,
then
we can say that the cosmological constant in our world is necessary
to get a gauge symmetric theory in our universe 
even if
how small it is.
%
%
The numerical analysis is shown by the curve A in Fig.\ref{mMfig}, 
from which we can assure
the statement given above.

\vspace{.5cm}
\noindent{\bf Solution(I-2)}
Next, we consider the solution {\bf (I-2)}. For this background,
$u$ is obtained in the same form given for the background {\bf (I-2)} as
\beq
 u(z)=c_1Y^{-id} {}_2F_1(b_1,b_2;\bar{c};Y)
+c_2Y^{id}{}_2F_1(b_1',b_2';\bar{c}';Y), 
        \label{solds1}
\eeq
where $c_{1}$ and $c_{2}$ are constants of integration. The same notations are
used for the parameters in the solution, but $X$ is replaced by
\beq
 Y={1\over\rm{cosh}^2(\sqrt{\lambda}z)},   \label{para11}
\eeq
and $\hat{a}$ is changed to
\beq
  \hat{a}=-4M^2/\mu_d^2+3 .
\eeq
For this solution, the normalizability condition is also satisfied if
we choose $c_2=0$ as in the previous case. The boundary condition at
$z=z_0$ can be written by the same formula (\ref{boundzero00}) in terms of 
the first term of the solution given in (\ref{solds1}). 
Then we perform the similar analysis
as in the case of {\bf (I-1)} to find the eigenvalue $m^2$.

In this case, we can examine $m^2$ by expanding it near $M^2=0$ as
$m^2=m_0^2+\alpha M^2$ as above, and we find 
$\alpha=\langle u|A^2|u\rangle $ is positive
and $m^2=(1/2) M^2$ near $\lambda=0$ as in the case of solution {\bf (I-1)}.
So $m_0=0$ also in this case. The example of an explicit numerical evaluation 
is shown in Fig.\ref{mMfig} by the dashed curve. 
Then we can say that the gauge bosons 
are trapped on the brane also in this brane solution.
The important point is the existence
of a positive $\lambda$, and the sign of $\Lambda$ is not essencial.
This situation is
also seen in the graviton trapping.

\vspace{.7cm}

\noindent{\bf Solution(II)}
Finally, we consider the case of the background solution {\bf (II)}.
In this case the solution of the vector fluctuation 
(\ref{warp300}) can be written as
\bea
  &&\qquad\qquad\qquad\qquad
   u(z)=(1-y)^{-\tilde{c}}v(y)  \ ,          \label{adsol}
\\
 && v(z)=c_1\ {}_2F_1(\tilde{b}_1,\tilde{b}_2;1/2;y)
        +c_2\ y^{1/2}\ {}_2F_1(\tilde{b}_1+1/2,\tilde{b}_2+1/2;3/2;y) \ , 
        \label{solds2}
\\
  &&\qquad\qquad
       y=\sin^2(Hz) \ ,\qquad
      \tilde{c}=-{1\over 4}+{1\over 2}\sqrt{1+M^2/\mu^2} \ ,
\\
  &&\qquad
  \tilde{b}_1=-\tilde{c}-{1\over 4}\sqrt{1+4m^2/H^2}, 
   \quad \tilde{b}_2=-\tilde{c}+{1\over 4}\sqrt{1+4m^2/H^2}, \label{para22}
\eea
where $c_1$ and $c_2$ are constants of integration.

\vspace{.2cm}
In this case, the potential $V(z)$ diverges at $z_1=\pi/(2H)$. Then we need
a new boundary condition at $z_1$
\beq
  u(z_1)=0 \label{new}
\eeq
in addition to the condition (\ref{boundzero00}) at $z=z_0$.
We solve these two
conditions as follows. The condition (\ref{boundzero00}) is used to fix the 
ratio $c_2/c_1$ as
\beq
 c_2/c_1=-{B_0F(y_0)+F'(y_0)\over B_0G(y_0)+G'(y_0)} \label{bound21}
\eeq
where $F(y)={}_2F_1(\tilde{b}_1,\tilde{b}_2;1/2;y)$, 
$G(y)=y^{1/2}{}_2F_1(\tilde{b}_1+1/2,\tilde{b}_2+1/2;3/2;y)$
, $y_0=\sin^2(Hz_0)$ and $'=\partial/
\partial y$. The factor $B_0$ is expressed by
\beq
 B_0={1\over 2}({\mu\over H})^2\sqrt{1+{M^2\over \mu^2}} \ ,\label{b000}
\eeq

\begin{figure}[htbp]
\begin{center}
\voffset=15cm
  \includegraphics[width=8cm,height=6cm]{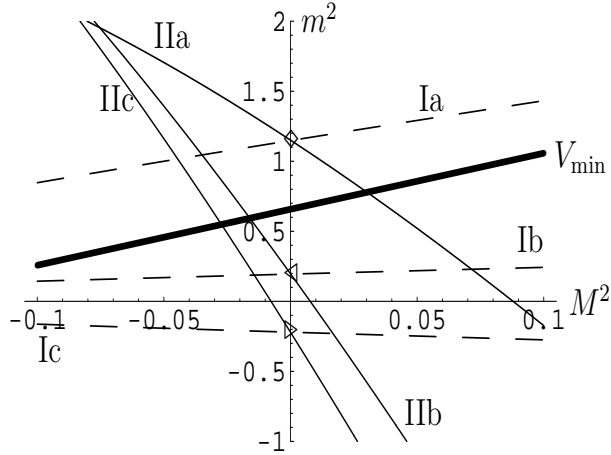}
\caption{The dashed curves show the $m^2$ versus $M^2$ curves
for $c=0$, Ia, Ib and Ic for  $H=0.5,~0.41,~0.4$, respectively. 
The curves for $c=\hat{c}$ are shown by
the solid lines  IIa, IIb and IIc  for $H=0.5,~0.41,~0.4$, respectively.
The points at $M=0$ for each lines are denoted by 
$\diamond,~\triangleleft,~\triangleright$. 
The curve $V_{\textrm{\scriptsize min}}$ is written for $H=0.5$.
In all cases, 
$m^2$ and $V_{\textrm{\scriptsize min}}$ are scaled by $H^2$,
where $\mu=0.55$. \label{curvemM}}
\end{center}
\end{figure}
\begin{figure}[htbp]
\begin{center}
\voffset=15cm
  \includegraphics[width=8cm,height=6cm]{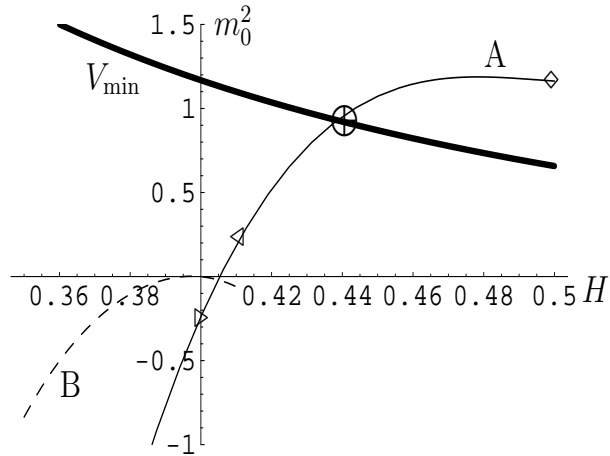}
\caption{The curves A and B show the eigenvalues $m_0^2$ versus $H$ 
for the parameters $\mu=0.55,~0.422$, respectively.
Both lines have end in right hand side to satisfy  $H<\mu$.
Also the points corresponding in Fig.{\ref{curvemM}}
are shown by $\diamond,~\triangleleft,~\triangleright$. 
The point for $m_0^2=V_{\textrm{\scriptsize min}}$
is denoted by $\bigoplus$.
$m_0^2=0$ is realized for $H=0.405$.
$V_{\textrm{\scriptsize min}}$ is written for $\mu=0.55$.
In all cases, $m_0^2$ and $V_{\textrm{\scriptsize min}}$ 
are scaled by $H^2$.
     \label{curvemh}}
\end{center}
\end{figure}

\begin{figure}[htbp]
\begin{center}
\voffset=15cm
  \includegraphics[width=6cm,height=4cm]{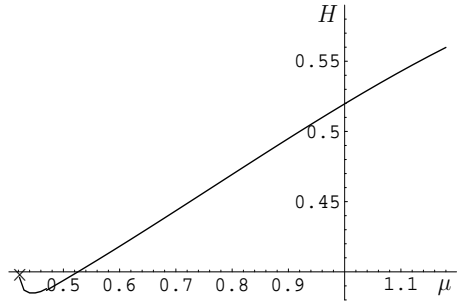}
\caption{$H$-$\mu$ curve with $m_0=M=0$.
The smallest value for $H$ is 0.385 
and the smallest value for $\mu$ is 0.422.
The solution does not exist for $\mu<0.422$. 
 \label{curvemuh}}
\end{center}
\end{figure}
 
\begin{figure}[htbp]
\begin{center}
\voffset=15cm
  \includegraphics[width=8cm,height=7cm]{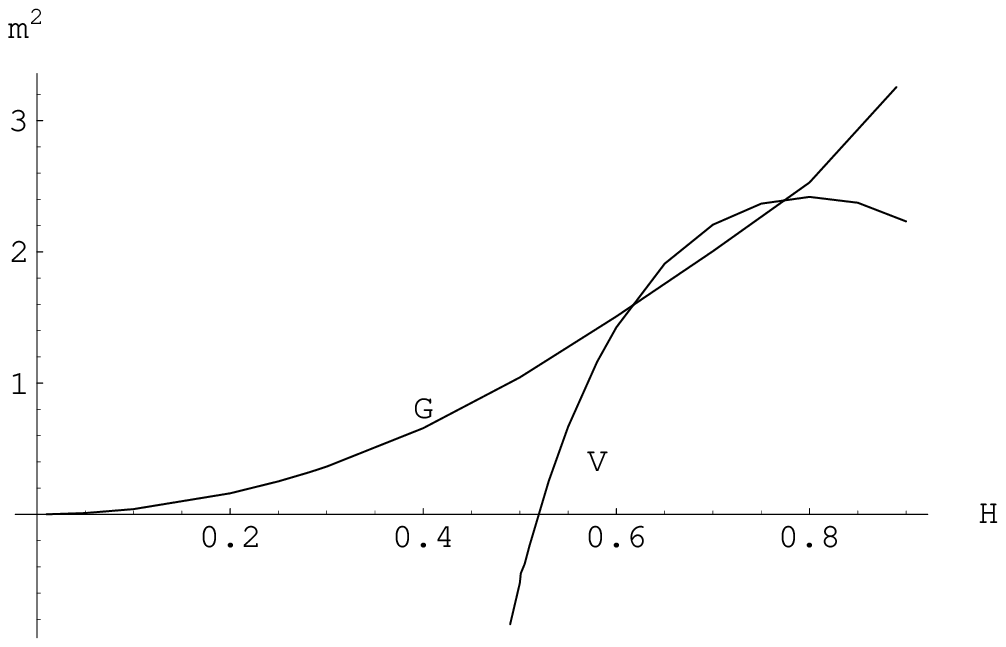}
\caption{The value $m$ satisfied the boundary conditions are shown
for the graviton (G) and gauge boson (V) versus $H$ for $\mu=1$.
\label{gvfig}}
\end{center}
\end{figure}

The condition (\ref{new}) can be written as,
\beq
  c_2/c_1=-2{\Gamma(1-\tilde{b}_1)\Gamma(1-\tilde{b}_2)\over
            \Gamma(1/2-\tilde{b}_1)\Gamma(1/2-\tilde{b}_2)}. \label{bound22}
\eeq
Using above two conditions (\ref{bound21}) and (\ref{bound22}), we can
obtain the eigenvalue $m^2$ as a value dependent on $M^2$ and other
parameters.
When we express
the $M^2$ dependence of $m^2$ as the form
$m^2=m_0^2+\alpha M^2$,
we find the following,
\beq
    \alpha=\langle u|A^2|u\rangle
           +{\partial u(z_1)\over \partial z}
            {\partial u(z_1)\over \partial M^2} \ .
\eeq
Then $\alpha$ is not necessarily positive. 
The numerical results are shown by the dashed lines in Fig.\ref{curvemM}.
In this case, $m_0^2$ depends on $H$
and becomes zero for specific $H$.
Furthermore $H$ has the upper bound so as to be 
$m_0^2<V_{\textrm{\scriptsize min}}$, 
which is condition to be trapped at $M=0$.
The relation between $m_0^2$ and $H$ is shown in Fig.\ref{curvemh}.
The upper bound to $H$ is the value for the point shown by $\bigoplus$
in the large $\mu$ region corresponding to curve A.
When $\mu$ takes a small value below the value corresponding to B,
the gauge field, that is, the fluctuation with $m_0=0$ 
is not trapped for $M=0$.
So $\mu$ has the lower bound for the gauge field to be trapped with $M=0$. 
The curve which  $H$ and $\mu$ draw 
for $m_0=M=0$ is shown in Fig.\ref{curvemuh}. 
$H$ also has the lower bound.
This means that the bulk gauge field is not trapped 
as gauge field for very small $H$
not only in the case of $\lambda=0$ as previously known. 

It should be astonishing that the gauge bosons can be trapped for
$\lambda<0$ differently from the case of the graviton which can not be
trapped in this case. To make clear the difference of gauge bosons and the
graviton, we show here the lowest mass of the trapped modes for both particles
in Fig.\ref{gvfig}. 
For graviton, $m^2$ is not able to become zero 
since $H\neq 0$.
This behaviour is also shown in \cite{bre}. While $m^2$ of gauge
boson crosses zero at finite $H$ as shown above for different parameters.
So there is a point in the parameter-space where gauge bosons are trapped.

\section{Vector localization: the case of $c=\hat{c}$}

Here we examine the vector localization for $c=\hat{c}$ case.
As mentioned above, the analysis is similar to the case of $c=0$
in the previous section.
Here the field equation of $A_M$ is given as 
\beq
 {1\over\sqrt{-G}}\partial_A[\sqrt{-G}
  (G^{AB}G^{CD}-G^{AC}G^{BD})\partial_B A_C] -
           [M^2 + \hat{c}\delta(y)]G^{DB}A_B =0 . \label{eq1}
\eeq
The $\delta$ function term appears newly.
Because of this term, the identity is presented by
\beq
 \partial_A([M^2 + \hat{c}\delta(y)]\sqrt{-G}G^{AB}A_B)=0 .  \label{identity}
\eeq
However the equation (\ref{identity}) can reduce to Eq.(\ref{identity300}) as
follows. Consider near $y=0$ of the equation (\ref{eq1}) for $D=y$, and 
integrate it over $y$ for the region $-\epsilon<y<\epsilon$, where $\epsilon$
is an infinitesimally small number. Then we find
$A_y(x,0)=0$ which implies $A_y(x,y)$ is odd with respect to the reflection
$y\to -y$. Then we can write (\ref{identity}) as,
\beq
 M^2\partial_A(\sqrt{-G}G^{AB}A_B) 
    + \hat{c}\delta(y)\partial_{\mu}(\sqrt{-G}G^{\mu\nu}A_{\nu})=0 .
     \label{identity2}
\eeq
Integrating this again in the region $-\epsilon<y<\epsilon$, we find 
$\partial_{\mu}(\sqrt{-G}G^{\mu\nu}A_{\nu})|_{y=0}=0$. As a result, the
identity (\ref{identity}) can be written as
\beq
  \partial_A(\sqrt{-G}G^{AB}A_B)=0, \label{identity3} 
\eeq
at any value of $y$. 
So we can solve the above field equation by using (\ref{identity3}).
The expansion of the field equation in terms of 
four-dimensional mass eigenstates leads only to
replace $M^2$ with $M^2+c\delta(y)$ in Eqs.(\ref{tra00})-(\ref{transv-t}).

Again we need only to consider the equation for $A_i^T$.
Then the Schr\"{o}dinger type equation (\ref{warp300})
has the potential
\beq
 V(z)={1\over 4}(A')^2+{1\over 2}AA''+A^2M^2+\hat{c}\delta(|z|-z_0) 
\eeq
where $\Lambda<0$ is considered by the definition of $\hat{c}$.
The explicit potentials are given as follows,

{\bf (I-1) Solution for $\lambda>0$ and $\Lambda<0$ :}
\beq
 V(z)= {1\over 4}\lambda [{4M^2/\mu^2+3\over\rm{sinh}^2(\sqrt{\lambda}|z|)}+1]
      -({\kappa^2\tau\over 6}-\hat{c})\delta(|z|-z_0) , \label{pot3}
\eeq

{\bf (II) Solution for $\lambda<0$ and $\Lambda<0$ :}
\beq
  V(z)= {1\over 4}H^2\left[{4M^2/\mu^2+3\over\cos^2(H z)}-1\right]
      -({\kappa^2\tau\over 6}-\hat{c})\delta(|z|-z_0) \ , \label{pot4}
\eeq
Away from the $\delta$-function term,
the bound state solutions are given by Eqs.(\ref{solb00}) and (\ref{adsol})
respectively.
On the other hand, the boundary condition at $z=z_0$ is changed as
\beq
{du(z_0)\over dz}=-{1\over 2}({\kappa^2\tau\over 6}-\hat{c})u(z_0).
    \label{boundzero}
\eeq
The results of our analyses are summarized as follows for the above two
solutions.

\vspace{.5cm}
\noindent {\bf For Solution (I-1):}
In this case
the consideration on the normalizability yields the only loose condition
$\m^2<\lambda/4$ as in the case of $c=0$.
However the  
relation between $m^2$ and $M^2$ is noticeably different.  
For this case with the brane coupling, $\alpha$ in Eq.(\ref{alph})
is negative.  
This is also confirmed both in numerically and analytically. 

\vspace{.2cm}
\noindent From the general representation (\ref{masseig3}),
$\alpha$ is expressed for small $M^2$ as
\beq
  \alpha=\partial_{M^2}m^2|_{M^2=0}=\langle u|A^2|u\rangle -{1\over\mu}|u(z_0)|^2,
        \label{massei4}
\eeq
where the second term in (\ref{massei4}) is coming from the term
$\hat{c}$ in the potential (\ref{pot3}). 
So we expect that $\alpha$ might be negative.
In fact when we examine $m^2$ near small $\lambda$,
the following relation is obtained, 
\beq
 m^2=-{\lambda\over 4\mu^2}M^2 \ .  \label{massei6}
\eeq
This implies that the massive bulk vector is trapped as a 
tachyonic vector on the brane when $\lambda>0$ differently from the case of
$\lambda=0$. Then this brane solution is unstable. It would be possible to
improve this instability by considering a dynamical scenario
that the trapped tachyonic vector would condense to change the brane-tension
$\tau$ to a small value until arriving at the stable point $\lambda=0$. 
The $\lambda$-dependence in Eq.(\ref{massei6}) seems to 
support this scenario.
So we can say that the case of $c=\hat{c}$ gives a
brane where the cosmological constant is always zero due to the 
gauge invariance on the brane. 
The numerical analysis is shown by the curve B in Fig.\ref{mMfig}.

\vspace{.3cm}

\noindent {\bf For Solution (II):}
In this case, the factor $B_0$ in Eq.(\ref{bound21})
is given instead of Eq.(\ref{b000}) as
\beq
 B_0={1\over 2}({\mu\over H})^2\left(\sqrt{1+{M^2\over \mu^2}}+
    {\sqrt{1+{M^2\over \mu^2}}-1\over \sqrt{1-{H^2\over \mu^2}}}\right) \ ,
\eeq
where
$H<\mu$ since $\mu^2-H^2=\kappa^4\tau^2/36$. 
Then $\alpha$ is obtained as
\bea
   \alpha=\langle u|A^2|u\rangle       
      +{\partial u(z_1)\over \partial z}
            {\partial u(z_1)\over \partial M^2} 
          -{1\over\mu}|u(z_0)|^2 \ .
\eea
The numerical estimation is shown by solid lines in Fig.\ref{curvemM}.
This implies also that massive bulk vector is 
trapped as a tachyonic vector on the brane.

\section{Summary}

We have found the possibility of the gauge fields trapping on the brane
through the analyses given here for vector fields. We have examined 
the localization of the fields on both dS and AdS
brane or on positive and negative $\lambda$. 
The bulk cosmological constant $\Lambda$ 
is also considered for both negative and positive cases. The model
considered here include the coupling of vectors and the brane
through the mass term of the vector with its coupling $c$.

\vspace{.3cm}
For the case of $c=0$, the brane model is normal and includes no
special brane coupling, the bulk gauge bosons 
can be trapped on the dS brane, i.e. for the brane of positive $\lambda$,
for both cases of positive and 
negative $\Lambda$. Then, we can say that the 
gauge symmetries in the bulk theory are preserved 
also on the dS brane. This situation 
is similar to the case of the graviton localization. Namely
the general coordinate invariance in the bulk is also preserved on the 
dS brane due to the graviton trapping.

The difference between the graviton and the gauge field
appears for $\lambda=0$, the RS brane. Actually, the gauge bosons 
can not be trapped on the RS brane since the cosmological constant 
is zero. Then we needed an appropriate coupling $c$
of gauge bosons and the brane for its trapping on the $\lambda=0$ brane
as shown above. 

\vspace{.2cm}
For non-zero $c$, the bulk mass of the vector field 
was needed to trap the vector fields on the brane as a gauge boson, i.e.
as a massless 4d vector. 
When this model is applied to the case of dS brane ($\lambda>0$),
we find an unexpected phenomena that
the bulk massive vectors are trapped on the dS brane
as tachyonic vectors. Then the dS brane is unstable in this model.


In this case
however, the bulk gauge symmetries is broken in spite of
the gauge symmetries on the brane. 
One possible   
idea to change this situation is to consider a model which contains the 
dilaton. And our effective action of massive vector 
fields can be obtained as an effective action by an appropriate field transformation \cite{Tati}
from the original gauge invariant theory.
If this is true, then the trapping of the gauge bosons will be solved
in terms of the the dilaton theory with bulk gauge symmetry.
On this point, we will discuss in a separate paper.


\vspace{.5cm}
Next, we summarize for the AdS brane ($\lambda<0$). 
In this case, the graviton
can not be trapped as a massless mode on the brane.
However we find that the 
bulk-vector bosons can be trapped on the brane as gauge bosons
for both models $c=0$ and $c=\hat{c}$. This is realized for
a specific value of $|\lambda|$, then we need a fine-tuning.

\vspace{.4cm}
Finally, we comment on the gap or the lower bound
for the continuum KK mode for the dS brane. 
It starts from $m^2 = \lambda/4$.
As a result, massive vector can be trapped in the range,
$m^2 < \lambda/4$, and its value is related to the bulk mass $M^2$ 
as $m^2=\alpha M^2$ where $\alpha$ is positive and dependent on
the parameters of the theory. We can expect that 
these localized fields of small mass vectors
would play some role in the cosmological scenario if they were really
exist. 


\end{document}